\documentclass[runningheads]{llncs}
\usepackage[T1]{fontenc}
\usepackage{graphicx}
\begin{document}

\title{Predicting Genetic Mutations from Single-Cell Bone Marrow Images in Acute Myeloid Leukemia Using Noise-Robust Deep Learning Models}

\author{
Garima Jain\inst{1} \and
Ravi Kant Gupta\inst{2} \and
Priyansh Jain\inst{2} \and
Abhijeet Patil\inst{2} \and
Ardhendu Sekhar\inst{2} \and
Gajendra Smeeta\inst{3} \and
Sanghamitra Pati\inst{4} \and
Amit Sethi\inst{2} \and
}

\authorrunning{G. Jain et al.}

\institute{
Indian Council of Medical Research-National Institute for Digital Health Research, Ansari Nagar, New Delhi, Delhi 110029 \and
Indian Institute of Technology, Bombay, Powai, Mumbai, Maharashtra 400076 \and
All India Institute of Medical sciences, New Delhi, Delhi-110029 \and 
Indian Council of Medical Research, Ansari Nagar, New Delhi, Delhi 110029
}

\maketitle

\begin{abstract}
In this study, we propose a robust methodology for identification of myeloid blasts followed by prediction of genetic mutation in single-cell images of blasts, tackling challenges associated with label accuracy and data noise. We trained an initial binary classifier to distinguish between leukemic (blasts) and non-leukemic cells images, achieving 90\% accuracy. To evaluate the model’s generalization, we applied this model to a separate large unlabeled dataset and validated the predictions with two haemato-pathologists, finding an approximate error rate of 20\% in the leukemic and non-leukemic labels. Assuming this level of label noise, we further trained a four-class model on images predicted as blasts to classify specific mutations. The mutation labels were known for only a bag of cell im- ages extracted from a single slide. Despite the tumor label noise, our mutation classification model achieved 85\% accuracy across four mutation classes, demonstrating resilience to label inconsistencies. This study highlights the capability of machine learning models to work with noisy labels effectively while providing accurate, clinically relevant mutation predictions, which is promising for diagnostic applications in areas such as haemato-pathology.
\end{abstract}

\section{Introduction}
Acute myeloid leukaemia (AML) is a highly aggressive cancer affecting the hematopoietic system. The understanding of the mutational landscape of AML has expanded significantly in recent years, leading to updates in the classification of AML and revolutionizing its prognostic stratification, treatment approaches, and response evaluation~\cite{jain2020next}. The primary curative approach involves intensive induction chemotherapy, followed by consolidation therapy or allogeneic hematopoietic stem cell transplantation\cite{cd34_garima}. Accurate risk assessment, which is crucial for selecting the most appropriate treatment, is currently based on genetic profiling of mutations and chromosomal abnormalities~\cite{kockwelp2024deep}, which also have proven to be the most predictive of overall survival and relapse-free survival~\cite{kurzer2023updates}. However, obtaining genetic information can take several days to weeks after diagnosis, incur substantial costs for healthcare as they are not universally available and require expert domain knowledge~\cite{dohner2022diagnosis}. \cite{peripheral_garima}

AML classification systems incorporate recurrent cytogenetic abnormalities, such as RUNX1:RUNX1T1, PML: RARA, MYH11: CBFB, and mutations in key genes like NPM1, FLT3, and CEBPA into the classification and prognostic models of AML\cite{hotspots_garima}. In recent years, new therapeutic agents have also been approved to complement intensive first-line treatments for specific genetic subgroups.

Cytomorphology may, in some cases, also lead to the suspicion of possible underlying genetic mutations. However, to the best of our knowledge, while machine learning has been leveraged to identify blasts from single cell images or to differentiate between blasts of acute myeloid leukemia and acute lymphoid leukemia, only a couple of studies \cite{hehr2023explainable} and \cite{dohner2022diagnosis} have leveraged deep learning to identify underlying genetic mutations on haemato-pathology smears.

In this study, we employ a multi-step deep learning method for automated segmentation and identification of myeloid blasts using the Giemsa-stained images of hu- man bone marrow samples. The downstream pipeline then identifies status of NPM1, RUNX1:RUNX1T1, and CBFB: MYH11. mutations between samples of Acute Myeloid Leukemia (AML) and healthy controls and predict the mutation.

\begin{figure}
    \centering
    \includegraphics[width=1\linewidth]{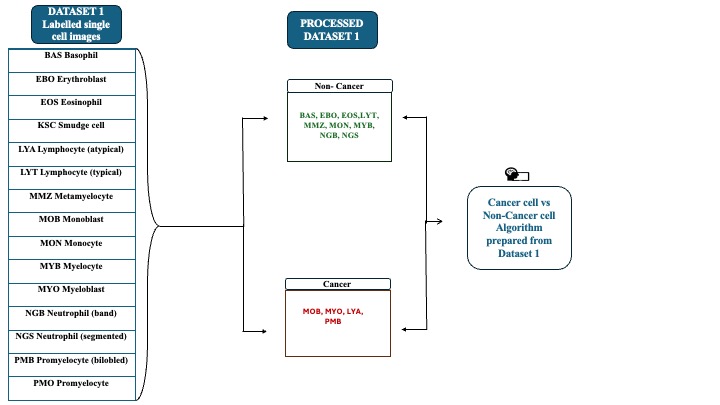}
    \caption{Our methodology comprises two key phases: (1) cancer detection through a threshold-based binary classification model for distinguishing malignant from benign cells}
    \label{fig1}
\end{figure}

\section{Methodologies}
\begin{figure}
    \centering
    \includegraphics[width=1\linewidth]{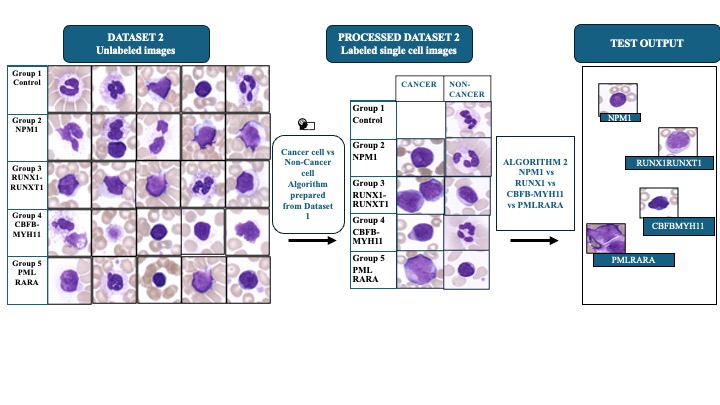}
    \caption{Mutation classification via a multi-class deep learning framework for subtype identification of specific cancer mutations using high-resolution single-cell images}
    \label{fig2}
\end{figure}

Our methodology comprises two key phases: (1) cancer detection through a threshold-based binary classification model for distinguishing malignant from benign cells shown in Figure \ref{fig1}, and (2) mutation classification via a multi-class deep learning framework for subtype identification of specific cancer mutations using high-resolution single-cell images shown in Figure \ref{fig2}. Advanced preprocessing, data augmentation, and class balancing techniques were implemented to mitigate is- sues with noisy labels, data sparsity, and class imbalances, thereby enhancing model robustness, sensitivity, and reliability\cite{her2_garima}. This integrated, multi-stage approach provides a comprehensive framework for cellular-level diagnostics, enabling accurate cancer detection and mutation-specific profiling within complex biological datasets.

\subsection{Datasets}
The experiments utilized two distinct datasets of acute myeloid leukemia (AML).

\noindent \textbf{Dataset 1 (D1):} This dataset, AML-Cytomorphology MLL Helmholtz, contains over 170,000 de-identified, expert-annotated cells from bone marrow smears of 945 patients. The dataset comprises a total of twenty-one classes, including two cancerous, sixteen non-cancerous, and three irrelevant classes as determined by domain experts. The cells were stained using the May-Gru¨nwald-Giemsa/Pappenheim stain and acquired via brightfield microscopy at 40x magnification. This dataset serves as a vital resource for developing and validating computational methods in diagnostic medicine, particularly for hematologic malignancies.

\noindent \textbf{Dataset 2 (D2):} The second dataset, AML- Cytomorphology MLL Helmholtz, includes data on four common subtypes of acute myeloid leukemia (AML) characterized by specific genetic anomalies and typical morphological features as classified by the WHO in 2022. The subtypes are: (i) APL characterized by the PML: RARA fusion, (ii) AML with a mutation in NPM1, (iii) AML marked by the CBFB: MYH11 fusion (excluding NPM1 mutation), and (iv) AML with the RUNX1:RUNX1T1 fusion. Addition- ally, the dataset includes a control group composed of healthy stem cell donors. A total of 189 peripheral blood smears were digitized, each containing between 99 and 500 white blood cells, scanned at 40x magnification. This dataset captures the morphological complexity of acute myeloid leukemia and provides a substantial basis for mutation classification.

\subsection{Algorithm}
\textbf{Cancer Detection -- Data Preprocessing and Model Development}: A notable challenge in processing Dataset 1 (D1) was the pronounced class imbalance between cancerous and non- cancerous cell types. Simple random sampling would have resulted in significant under-representation of cancerous cells, which could bias the model toward the prevalent non- cancerous class. To address this, we implemented a stratified proportional sampling algorithm, which ensured balanced representation across both training and validation sets, thus enhancing model generalization and reducing bias by providing equivalent exposure to both classes.

The cancer detection model was based on ResNet-18\cite{resnet}\ architecture that was pretrained on ImageNet\cite{imagenet}, to classify individual single-cell images as cancerous or non-cancerous. To reduce the impact of classification errors, a 5\% error threshold was established; if fewer than 5\% of the cells for a given patient were classified as cancerous, the patient was labeled as non-cancerous. This thresholding approach minimized the likelihood of misclassifying patients with sparse cancerous cells and improved classification reliability.

\noindent \textbf{Mutation Detection:} Addressing Noisy Labels and Classification
Considering the potential for misclassification in which non-cancerous cells may be incorrectly identified as cancerous, we approached this issue as a noisy label problem. Smooth cross-entropy\cite{szegedy2016rethinking} was employed to mitigate the impact of noisy labels, using a noisy label percentage of 20\%. This method allowed for more flexibility in learning from potentially mislabeled data, reducing the impact of erroneous labels on model performance. A multi-class classification model was developed to classify cancerous cells into one of four mutation types.

\section{Results}
We utilized a pretrained ResNet-34\cite{resnet} as our primary feature extractor, as it demonstrated superior performance compared to other architectures. For the ablation study, we experimented with various feature extractors, including ResNet-18\cite{resnet}, ResNet- 50, VGG13, VGG16, VGG19\cite{vggnet}, ConvNeXt Tiny\cite{convnext}, Swin Transformer Tiny\cite{swin}, and EfficientNet\cite{efficientnet}. Our findings indicate that ResNet-34\cite{resnet} provided the most effective feature extraction capabilities, justifying its selection for our approach. Dataset one was partitioned into an 80-20 split for training and validation, with optimization performed using the Adam optimizer (learning rate 0.001, weight decay 0.001, batch size 16). Regularization was reinforced through a dropout layer (0.3) following feature extraction, enhancing model robustness, and mitigating overfitting. Early stopping, with a patience level of 50 epochs and an error tolerance (delta) of 0.02, ensured efficient convergence. Specifically, the ResNet-18 model was trained in two configurations: first with a frozen encoder, training only the classification head, achieving maximum training and validation accuracies of 0.789 and 0.837, respectively, and then with full encoder fine-tuning, resulting in improved maximum accuracies of 0.918 for training and 0.892 for validation. All encoders were initialized with ImageNet\cite{imagenet} pre-trained weights, facilitating effective transfer learning for cancer detection.

For mutation detection, the same encoder architectures were applied to Dataset 2, which was split into 72\%, 18\%, and 10\% for training, validation, and testing, respectively. Optimization was conducted using the Stochastic Gradient Descent (SGD) optimizer, configured with a learning rate of 0.001, weight decay of 0.001, momentum of 0.9, and a batch size of sixty-four to support stable convergence. A dropout layer (0.3) was incorporated post-feature extraction and preceding the classification head to enhance generalization and minimize overfitting risks. Early stopping, with a patience of 50 epochs and an error delta of 0.02, further improved training efficiency by halting training upon stabilized convergence, ensuring computational efficiency and model performance reliability.

The performance evaluation of the cancer detection and mutation classification models described in Table 1, was conducted using several rigorous metrics, including accuracy, precision, recall, and F1-score. The model achieved an ac- curacy of approximately 90\% in distinguishing cancerous from non-cancerous cells, demonstrating its robustness in high-stakes diagnostic tasks. The cancer detection model achieved an accuracy of approximately 90\% in classifying single cell images of cells from Bone marrow into cancerous and non-cancerous categories for D1. This performance metric indicates a high reliability in differentiating malignant from benign cells, further underscored by precision and re- call values that capture the model’s capability to correctly identify positive cases while minimizing false positives and negatives. The F1-score offers a balanced measure, providing insights into the model’s robustness in practical, clinical scenarios. By applying a stringent 5\% error threshold, the model effectively reduced false positives in the classification of non-cancerous samples. This threshold ensured that the model maintained a high specificity, thereby minimizing the likelihood of incorrectly identifying healthy cells as can- cerous. This approach enhances the model’s reliability for clinical applications, where reducing false positives is critical to prevent unnecessary diagnostic follow-ups and treatments.

The mutation classification model achieved an 85\% accuracy across the four classes using the smooth cross entropy loss that is robust to data mislabeling. This result was obtained by employing a majority voting mechanism based on aggregated predictions from individual single-cell images to enhance classification robustness across specific cancer mutation categories. This approach allowed for a more reliable consensus on mutation classification, reducing variability and aligning closely with expert-annotated benchmarks. The high concordance between the model’s predictions and expert analyses highlights the model’s potential in supporting accurate and scalable mutation classification.
Collectively, these results underscore the efficacy of the proposed methodologies in advancing diagnostic precision in cancer detection and mutation typing, establishing a robust framework for further exploration and deployment within clinical diagnostics.

\section{Discussion}
Our study demonstrates robust performance in identifying clinically relevant mutations in AML through cytomorphology images, aligning with advancements in the field while offering distinct improvements in data handling and accuracy. For instance\cite{kockwelp2024deep}, developed a pipeline using Pappenheim- stained bone marrow smears to predict genetic mutations such as NPM1, achieving an AUROC of 0.86 for NPM1 mutations. Similarly employing deep learning on cytomorphology, our approach utilizes a multi-step process with CNNs and noise-resistant loss functions, achieving comparable accuracy (85\%) even under noisy label conditions. Other models, such as the Morphogo system\cite{lv2023high} trained on a massive dataset of bone marrow cells (2.8 million images), report high overall accuracy in morphology-based classification, though they often emphasize cell-type identification over mutation prediction. Meanwhile, \cite{boldu2021deep} developed AL- NET, which focused on lineage-specific classification among leukemia subtypes, with high sensitivity and specificity in differentiating acute promyelocytic leukemia and myeloid leukemia lineages. While these models show valuable results, our approach uniquely combines mutation-specific classification with label noise handling, enhancing model robustness in noisy, real-world clinical data contexts and underscoring its potential for broader clinical application.

Our study aligns with and extends prior work on AML subtype identification using deep learning while offering unique insights into handling label noise and robust mutation detection. SCEMILA,\cite{hehr2023explainable} an explainable AI model, employed attention mechanisms to identify diagnostically relevant cells in AML, achieving F1 scores as high as 0.86 for PML: RARA subtypes with agreement from human expert annotations. Similarly, hierarchical CNN approach for APL detection gave high precision and recall in a previous
 
Study\cite{kockwelp2024deep}, demonstrating the utility of CNNs in rare AML subtype detection, especially in low-resource settings without immediate genetic testing. Our study achieved comparable performance metrics, with an 85\% accuracy for mutation classification, addressing similar diagnostic challenges but with a focus on resilience to label noise. By employing noise- handling techniques and smooth cross-entropy loss function, our model demonstrated stable performance under approximately 20\% noisy labels. Unlike SCEMILA and Eckardt’s models, which focused on attention and cell-specific feature extraction, our approach provides broader applicability across multiple mutations (e.g., NPM1, RUNX1:RUNX1T1) in noise-prone environments. This focus on robust classification amidst label inconsistencies enhances the model’s potential for clinical integration, allowing for reliable mutation prediction without extensive label refinement, a common barrier in AML cytomorphology applications.

\section{Conclusion}
Label noise is a prevalent challenge in medical imaging, especially in tasks like cancer detection, where data labeling requires domain-specific expertise and is often prone to errors. To mitigate the effects of noisy labels, various strategies and models have been implemented to ensure robustness and improve diagnostic reliability in deep learning applications. Regularization methods, such as robust loss functions (e.g., cross-entropy with noise handling adjustments) and dropout layers, prevent overfitting to noisy data by focusing the model on reliable signal patterns. This method is particularly valuable in image classification where features vary significantly, and the goal is to avoid learning from anomalous data due to noise. Hybrid methods like semi-supervised learning use a mix of labeled and unlabeled data, where the latter often pro- vides cleaner representations for model training. Contrastive learning, by focusing on distinguishing between similar and dissimilar pairs, can help the model ignore noisy labels by emphasizing the underlying structure of images rather than la- bels, which is useful in cytomorphology and medical imaging with inherent noise. In complex medical datasets where certain image features may inherently lead to more frequent mis- labeling (e.g., in AML cell variations), instance-dependent noise modeling has proven effective. This approach tailors noise handling to specific data characteristics, thus providing an advantage in nuanced datasets like medical images of bone marrow samples, where noise distribution varies at the instance level.

\subsection*{Compliance With Ethical Standards}
This research study was conducted retrospectively using hu- man subject data made publicly available. Ethical approval was not required as confirmed by the license attached with the open access data.

\subsection*{Acknowledgments}
No funding was received for conducting this study. The authors have no relevant financial or non-financial interests to disclose.

\subsection*{Authorship Statement}
Garima Jain: Data collection, data analysis, writing the first draft of manuscript.
\newline
Ravi Kant Gupta: Discussion writing, Reviewing, and editing of the manuscript.
\newline
Priyansh Jain: Algorithm Development, Model training, testing
\newline
GJ, RKG and PJ contributed equally to this paper.
\newline
Abhijeet Patil: Statistical analysis, interpretation of the results
\newline
Ardhendu Shekhar: Statistical analysis, interpretation of the results
\newline
Gajendra Smeeta: Overall supervision and mentorship of the research, reviewing and finalizing the manuscript. 
\newline
Sanghamitra Pati: Conceptualization \& Design, framing aim and research objectives, Overall supervision, and mentorship of the research, reviewing and finalizing the manuscript.
\newline
Amit Sethi: Conceptualization, framing aim and research objectives, Overall supervision, and mentorship of the research, reviewing and finalizing the manuscript.
\newline
SP, AS are  corresponding authors.

\subsection*{Guarantor Information}
Garima Jain accepts full responsibility for the finished work and the conduct of the study, has access to the data, and controlled the decision to publish.

\subsection*{Ethics Statement} 
No ethical clearance was required as the study does not deal with human/animal data.

\subsection*{Funding statement} 
Not Applicable

\bibliographystyle{splncs04}
\bibliography{refs}

\end{document}